\documentclass[11pt]{article}
\usepackage{amsmath,amsfonts, amssymb, braket, bbold}
\usepackage{tikz}
\usetikzlibrary{trees,er,snakes,shapes,mindmap}
\usepackage{titlesec}
\titleformat{\section}
 {\normalfont\fontfamily{bch}\fontsize{12pt}{16pt}\bfseries\color{black}}
{\thesection}{1em}{}
\def \beq  {\begin{equation}}
\def \eeq  {\end{equation}}
\def \beqar {\begin{eqnarray}}
\def \eeqar {\end{eqnarray}}
\allowdisplaybreaks
\def\sqr#1#2{{\vcenter{\vbox{\hrule height.#2pt
\hbox{\vrule width.#2pt height#1pt \kern#1pt
\vrule width.#2pt}\hrule height.#2pt}}}}

\def\S {{\cal S}}

\def\vf {{\varphi}}

\def\Tr {{\rm Tr}}

\def\del {\partial}

\def\bz {{\bar{z}}}

\def\C {{\cal C}}
\def\D {{\cal D}}

\def\H {{\cal H}}

\def\N{{\cal N}}

\def\U{{\cal U}}

\def\vf {{\varphi}}

\mathchardef\mhyphen="2D
\begin{document}
\fontfamily{bch}\fontsize{12pt}{17pt}\selectfont
\def \CMP {{Commun. Math. Phys.}}
\def \PRL {{Phys. Rev. Lett.}}
\def \PL {{Phys. Lett.}}
\def \NPBProc {{Nucl. Phys. B (Proc. Suppl.)}}
\def \NP {{Nucl. Phys.}}
\def \RMP {{Rev. Mod. Phys.}}
\def \JGP {{J. Geom. Phys.}}
\def \CQG {{Class. Quant. Grav.}}
\def \MPL {{Mod. Phys. Lett.}}
\def \IJMP {{ Int. J. Mod. Phys.}}
\def \JHEP {{JHEP}}
\def \PR {{Phys. Rev.}}
\def \JMP {{J. Math. Phys.}}
\def \GRG{{Gen. Rel. Grav.}}
\begin{titlepage}
\null\vspace{-62pt} \pagestyle{empty}
\begin{center}
\vspace{1.3truein} {\Large\bfseries
A Note on Coadjoint Orbits for Multifermion Systems}\\
\vskip .15in
{\Large\bfseries ~}\\
\vskip .5in
{\Large\bfseries ~}\\
{\large\sc V.P. Nair}\\
\vskip .2in
{\itshape Physics Department,
City College of the CUNY\\
New York, NY 10031}\\
 \vskip .1in
\begin{tabular}{r l}
E-mail:&\!\!\!{\fontfamily{cmtt}\fontsize{11pt}{15pt}\selectfont vpnair@ccny.cuny.edu}\\
\end{tabular}
\vskip 1.5in
\centerline{\large\bf Abstract}
\end{center}
The coadjoint orbit action for a multifermion system, as an exact description of its dynamics, is considered.
A parametrization of the variables involved is given which facilitates the
approximation of this by another
coadjoint orbit action suitable for expansions around the Fermi surface,
recovering various actions which have been used in previous literature.
The presentation of this in terms of functions on phase space with
star-products as well as further truncations are briefly considered.
\end{titlepage}
\pagestyle{plain} \setcounter{page}{2}
\section{Introduction}
There is a long history and a large body of literature on the use of coadjoint orbits in physics going back to the 1970s. This is largely due to the fact that the Kostant-Kirillov-Souriau action on the coadjoint orbits of a 
group is the paradigmatic example of geometric quantization
\cite{reviews}.
They are particularly useful for physical problems 
involving nontrivial fiber bundles, such as
particle dynamics in the presence of a monopole, 
anomalous symmetries, particles with nonabelian charges, etc.
\cite{bal}.
They are also especially suited to describing the dynamics of
multifermion systems, 
something which was noted shortly 
after the initial work on geometric quantization.
In a series of papers, Rosensteel, Rowe and others formulated 
multifermion systems using coadjoint orbits \cite{rosen1},
see also the review \cite{rosen2}.
They were also able to include interactions within the Hartree-Fock approximation using orbits of the so-called determinantal type.
The action was given in terms of a density matrix, strictly speaking an occupancy matrix, 
although the group structure for the orbits was clear from the
general discussion. 
Quantum Hall systems naturally lead to a droplet of fermions
in phase space.
Classically, fluctuations of the droplet will correspond to an orbit of 
area-preserving diffeomorphisms, quantum theoretically to
a suitable orbit in the single-particle Hilbert space.
The corresponding dynamics in terms of coadjoint orbit actions
have been discussed by many authors \cite{{wadia},{sakita}}.
There is also a straightforward
generalization to higher dimensional Hall effect with both abelian
and nonabelian background magnetic fields \cite{KN1};
fluctuations of the background gauge fields can be accommodated as well
\cite{Kar}.
In this context, we may note that, by
virtue of projection to a given Landau level, the
manifold where we obtain the
droplet of fermions is in fact the phase space. 
Dynamics described
by the coadjoint orbit actions are then equivalent to
bosonization around the Fermi surface.
(Bosonization is a topic of longstanding interest with an enormous body of literature. Many of the papers touch upon coadjoint orbits as well \cite{{bosonization},{poly}}.)
More recently, bosonization around the Fermi surface has also been discussed
using orbits generated by exponentiation
of Lie derivatives in phase space \cite{son}.
This is somewhat similar to 
deformation quantization, although the deformation parameter is not
$\hbar$.

Placed in this context, the purpose of the present note is two fold.
First of all we point out that there are three levels of description
for a multifermion system. If ${\mathcal N}$ is the dimension of
the Hilbert space of states of a given multiparticle system,
then an exact quantum description of the dynamics 
is obtained using the orbits of the
type $SU({\cal N}) / U({\cal N} -1)$.
If we neglect intrinsic two-particle and multi-particle correlations,
and consider only the dynamics generated by unitary transformations on
the single-particle Hilbert space, we get a description using orbits
of the type $SU(N)/ [SU(K)\times SU(N-K)\times (U(1)]$, for a $K$-fermion system
where $N$ is the dimension of the single-particle Hilbert space.
This reduction is equivalent to considering orbits of the determinantal type
mentioned earlier and is essentially the Hartree-Fock approximation,
but our formulation will also indicate how one might attempt to go
beyond this approximation.
By use of coherent state wave functions, one can convert the resulting 
action to one in terms of classical functions and star-products, 
producing a description in terms of phase space densities.
The star-products lead to a series expansion in terms of derivatives
of the phase space density and the Hamiltonian.
While this is still formally exact, the truncation of the series at any finite order reduces it to something similar to a
semiclassical description. However, the expansion parameter is not
$\hbar$, rather it is an expansion involving inverse
powers of the symplectic form on the phase space.
The truncation of the series may be an acceptable
approximation at large $N$.
This is basically the procedure used for 
the quantum Hall systems  and for bosonization in
\cite{wadia}-\cite{bosonization} as well as in some recent work \cite{son}.
Secondly, by starting with an exact description and deriving the
approximations in stages, and enunciating the assumptions involved in such reductions, we hope to give a proper placement in context
to various approaches used in the
past.

In the next section we show how the exact dynamics of a general quantum system is described using the coadjoint orbit action.
We formulate this, as well as various approximations to follow,
in terms of finite-dimensional Hilbert spaces.
Equivalently, one can think of the relevant phase spaces
as compact with finite volume. We will assume that there is
a suitable limiting procedure for taking the limit
of infinite dimension for the Hilbert space, or infinite volume for
the phase space, if needed.
In section 3, we consider how this applies to multifermion systems.
A parametrization of the variables involved is given in a form suitable
for approximating the time-evolution to
unitary transformations on the single-particle Hilbert space.
It is shown how this leads to the Hartree-Fock approximation and also
recovers the method used for edge modes around the Fermi surface,
as in \cite{wadia}-\cite{Kar} and some of the papers in \cite{bosonization}.
We then show how we can re-express this in terms of functions on phase
space and associated star-products. The paper concludes with a short discussion.
\section{Coadjoint orbits for a general quantum system}
The dynamics of a quantum mechanical system can be described by a path-integral on a coadjoint orbit of the group of unitary transformations.
While this method can be used for any quantum system, it is particularly useful for  multifermion systems, where, by use of star-products and 
other ideas related to noncommutative geometry, the coadjoint orbit action can be simplified and
effective actions for excitations around the Fermi surface can be easily constructed.

We start by considering the coadjoint orbit description for a general quantum mechanical system. This part of our analysis is not restricted to multifermion systems.
Consider a Hilbert space $\H$ of states of dimension ${\cal N}$.
(We can eventually take ${\cal N} \rightarrow \infty$, if needed.)
A general state can be expanded as 
\beq
\ket{\psi} = \sum_\alpha c_\alpha \, \ket{\alpha} \equiv \ket{c}
\label{cod1}
\eeq 
where $\{ c_\alpha \}$ form a set of complex coefficients in the expansion
and $\ket{\alpha}$ form a basis of states.
We can view the set $\{ c_\alpha \}$ as a complex vector 
in $\mathbb{C}^{\cal N}$.
The normalization of the state imposes the condition
$\sum c^*_\alpha c_\alpha = 1$.
Further a common phase is irrelevant and can be removed, so $c_\alpha$ define
the complex projective space $\mathbb{CP}^{{\cal N}-1}$.
Each state is characterized by the $c$'s. So we can denote
$\ket{\psi}$ above as $\ket{c}$, as indicated in (\ref{cod1}).
Taking $\ket{\alpha}$'s to be an orthonormal set, we also have
\beq
\braket{\alpha|c} = c_\alpha, \hskip .3in
\braket{c|\beta} = c^*_\beta
\label{cod2}
\eeq
Assume we have a completeness relation for the states in the $c$-basis,
of the form
\beq
\int d\mu (c, c^*) \, \ket{c} \bra{c} = \mathbb{1}
\label{cod3}
\eeq
We can then write
\beq
\int d\mu (c, c^*) \, \braket{\alpha|c} \braket{c|\beta} 
= \braket{\alpha|\beta} = \delta_{\alpha \beta}
\label{cod4}
\eeq
Using (\ref{cod2}) for the left hand side, we find
\beq
\int d\mu (c, c^*) \, c_\alpha \, c^*_\beta = \delta_{\alpha\beta}
\label{cod5}
\eeq
This is obtained if $d\mu$ is the standard volume element for
$\mathbb{CP}^{{\cal N}-1}$. Going back to (\ref{cod3}), we 
then see that it holds as a completeness relation if we identify the
integration measure as the volume element for $\mathbb{CP}^{{\cal N}-1}$.

We now consider the time-evolution of the state $\ket{c}$. 
The infinitesimal time-evolution matrix element is given by
\beqar
\bra{c'} e^{- i H \epsilon} \ket{c} &\approx& \sum \bra{\beta} c'^*_\beta
\left[ 1- i H \epsilon \right] c_\alpha \ket{\alpha}
\nonumber\\
&=&\bra{\beta} \Bigl(c^*_\beta c_\alpha \ket{\alpha} + \epsilon {\dot c^*_\beta} c_\alpha \ket{\alpha}\Bigr) - i \epsilon \bra{\beta} H \ket{\alpha} c^*_\beta c_\alpha
\nonumber\\
&=& 1 + i \left( -i {\dot c^*_\alpha} c_\alpha - c^*_\beta H_{\beta\alpha}
c_\alpha \right) \epsilon\nonumber\\
&=& 1 + i \left( i {c^*_\alpha} {\dot c_\alpha} - c^*_\beta H_{\beta\alpha}
c_\alpha \right) \epsilon
= e^{i S (\epsilon )}
\label{cod6}
\eeqar
where we have used $\braket{\beta|\alpha} = \delta_{\beta\alpha}$,
$c^*_\alpha c_\alpha = 1$
and $H_{\beta \alpha} = \bra{\beta} H \ket{\alpha}$.
Also, as is usually done for
the path-integral, we have expanded $c'^*$ in terms of $c^*$, writing
$c'^* = c^* + \epsilon \, {\dot c}^*$.
The insertions of completeness relation for time-evolution over
a finite interval brings in factors of $d\mu$. Thus
we get the path integral
\beq
\bra{c', t} e^{ - i H t} \ket{c, 0} = \lim_{n \rightarrow \infty, \, \epsilon_i \rightarrow 0}
\int \prod_1^n d\mu (c^{(i)} , c^{(i)*})~ e^{ i \sum_{i=0}^n\, S(\epsilon_i )}
\label{cod7}
\eeq
where $t = \sum_{i=0}^n \epsilon_i$.
There is a path in $c$-space given by
$(c', c^{(n)}, c^{({n-1})}, \cdots, c^{(1)}, c)$. We integrate over all
intermediate points in (\ref{cod7}).
This result shows that the time-evolution for any system can be described by
a path-integral over the space of states, i.e., over $\mathbb{CP}^{{\cal N}-1}$.
The relevant action is given by
\beq
S = \int dt \left[ i c^*_\alpha {\dot c}_\alpha - 
c^*_\beta H_{\beta \alpha} c_\alpha
\right]
\label{cod8}
\eeq
Since the coefficients $c_\alpha$ obey the normalization condition and
we can remove a common phase, we can obtain them from the action of
a unitary transformation ${\cal U}_{\alpha\beta}$ on a particular vector $\{ c_\alpha\}$.
Explicitly
\beq
c_\alpha = {\mathcal U}_{\alpha 1} 
\label{cod9}
\eeq
Using this in (\ref{cod8}), we find that
\beqar
S &=& \int dt \,\left[ i \,{\mathcal U}^*_{\alpha 1} {\del {\mathcal U}_{\alpha 1} \over \del t}
- {\mathcal U}^*_{\beta 1} H_{\beta\alpha} {\mathcal U}_{\alpha 1} \right]\nonumber\\
&=&\int dt \,\Tr \left[\rho_0\left( i \,{\mathcal U}^\dagger {\del {\mathcal U} \over \del t}
- {\mathcal U}^\dagger H {\mathcal U}\right)\right] 
\label{cod10}
\eeqar
where $\rho_0 = \ket{1} \bra{1}$ is the projection to one of the states.
If we make the transformation
${\mathcal U} \rightarrow {\mathcal U} \, h$, $h \in U({\cal N} - 1)$, in the second line of this equation, the action is unchanged since
$h$ commutes with $\rho_0$ and hence
$S$ is the coadjoint orbit action defined on $SU({\cal N})/ U({\cal N}-1)$.
We can thus conclude that the time-evolution of any quantum state is described by suitable a coadjoint orbit action.
\section{Coadjoint orbits for a $K$-fermion system}
We can now apply or adapt this to the case of a $K$-fermion state
where the one-particle Hilbert space has
dimension $N$. We will allow for interparticle interactions as well.
If we have $N$ fermion creation/annihilation operators
$b^\dagger_\alpha$, $b_\alpha$, $\alpha = 1, 2, \cdots, N$,
the possible multiparticle states are of the form
$b^\dagger_{\alpha_1} b^\dagger_{\alpha_2}\cdots b^\dagger_{\alpha_K}
\ket{0}$ for $K = 1, 2, \cdots, N$. The total number of such states
is $2^N$. Consider the set of states for a fixed value of $K$,
i.e., a $K$-fermion system.  The number of possible states for the $K$-particle system 
is ${\cal N} = N!/(K! (N-K)! )$. The states 
$b^\dagger_{\alpha_1} b^\dagger_{\alpha_2}\cdots b^\dagger_{\alpha_K}
\ket{0}$ form a basis, they are the analogue of the states $\ket{\alpha}$ in
(\ref{cod1}).
 As in the single particle case, we can characterize the choice of a state by a complex number $c_A$, $A = 1, 2, \cdots, {\cal N}$,
with $c^*_A c_A = 1$,
corresponding to 
\beq
\ket{\psi} = \sum c_A \ket{A} \sim \sum c_{\alpha_1 \alpha_2 \cdots \alpha_K}  \, b^\dagger_{\alpha_1} b^\dagger_{\alpha_2}\cdots b^\dagger_{\alpha_K}
\ket{0}
\label{cod10b}
\eeq
The index $A$ is a composite index of the form
$\alpha_1 \alpha _2 \cdots \alpha_K$, antisymmetric under any exchange, so that
\beq
c_A \sim  c_{\alpha_1 \alpha _2 \cdots \alpha_K}
\label{cod31}
\eeq
A common phase for the $c$'s can be eliminated as well.
The action to be used in the path-integral is then 
\beq
\S = \int dt \left[ i c^*_A {\del c_A \over \del t} - c^*_A H_{AB} c_B
\right]
\label{cod30}
\eeq
We have allowed for the possibility of a general Hamiltonian with multiparticle
interactions.
There is an obvious $U({\cal N})$ symmetry in (\ref{cod30}),
corresponding to $c_A \rightarrow \U_{AB} c_B$, $\U \in U({\cal N})$.
Since we can remove an overall phase, this symmetry is reduced to
$SU({\cal N})$. The requirement $c^*_A c_A = 1$ breaks this
$SU({\cal N})$ symmetry to $U({\cal N} - 1)$. Therefore $c_A$ can be viewed
as parametrizing $\mathbb{CP}^{{\cal N} - 1} = SU({\cal N}/U({\cal N} - 1)$.
We may in fact write $c_A = \U_{A 1}$, similar to what was done
in (\ref{cod9}).
If we introduce local complex coordinates 
$z_\lambda$, $\lambda = 1, 2, \cdots, ({\cal N} -1)$,
for $\mathbb{CP}^{{\cal N} -1}$,
then $c_A = \U_{A 1}$ are given by
\beq
c_1 = \U_{11} = {1\over \sqrt{1+ \bz_\lambda z_\lambda}},
\hskip .2in
c_\lambda = \U_{\lambda 1} = {z_\lambda \over \sqrt{1+ \bz_\lambda z_\lambda}}
\label{cod30a}
\eeq

There is also a $U(N)$ action on the $c_A$'s given by
\beqar
c_{\alpha_1 \alpha_2 \cdots \alpha_K}
&\rightarrow& \Bigl(U_{\alpha_1 \beta_1}  U_{\alpha_2 \beta_2}  \cdots
U_{\alpha_K \beta_K} \Bigr) \, c_{\beta_1 \beta_2 \cdots \beta_K},
\hskip .2in U_{\alpha \beta} \in U(N)\nonumber\\
c_A &\rightarrow& g_{AB} c_B
\label{cod32}
\eeqar
$g_{AB}$ in the second line of this equation is an
${\cal N} \times {\cal N}$ matrix, corresponding to the 
irreducible representation of $U(N)$ obtained as the antisymmetric
product of $K$ fundamental representations.

The discussion so far is identical to what we did for any set of states as in 
(\ref{cod7}) and (\ref{cod10}), except for specializing
 the Hilbert space to correspond to $K$-fermion states. We want to show 
 how the action (\ref{cod30}) reduces to the form used in \cite{KN1}
 when we restrict the dynamics to unitary transformations of the
 single-particle Hilbert space, namely to transformations of the form
 (\ref{cod32}). Our purpose is also to show precisely what 
 approximations are being made in obtaining such a reduction.
To see this reduction explicitly, we need a parametrization of $c_A$
in terms of $SU(N)$ group elements, rather than
the one given in (\ref{cod30a}),
 which allows for an easy separation of the action of
 $U$'s, rather than the full transformation
 $\U \in SU({\cal N})$. 
 This can be done as follows.

As mentioned before, the condition $c^*_A c_A = 1$ breaks the symmetry 
$SU({\cal N})$ to $U({\cal N} -1)$. But this also entails the breaking of
$SU(N)$ to 
$H = SU(K) \times SU(N-K)$, as we explain below.
We can utilize this feature to find the parametrization
we are aiming for.
Towards this, we first decompose the basis states of the $K$-particle
Hilbert space (which transform as the fundamental representation of
$SU({\cal N})$) into irreducible representations of
$ SU(K) \times SU(N-K) $.
This is done by splitting the range of $\alpha$ labeling the fermion
operators $b_\alpha$, $b^\dagger_\alpha$ as
$i, j, \cdots = 1, 2, \cdots, K$,
$a, b, \cdots = 1, 2, \cdots, (N-K)$.
The $K$-particle states 
$b^\dagger_{\alpha_1} b^\dagger_{\alpha_2}\cdots b^\dagger_{\alpha_K}
\ket{0}$ then decompose as
\beq
b^\dagger_{\alpha_1} b^\dagger_{\alpha_2}\cdots b^\dagger_{\alpha_K}
\ket{0} = b^\dagger_{i_1} b^\dagger_{i_2}\cdots b^\dagger_{i_K}
\ket{0} \oplus 
b^\dagger_{a} b^\dagger_{i_2}\cdots b^\dagger_{i_K}
\ket{0} \oplus b^\dagger_{a_1} b^\dagger_{a_2} b^\dagger_{i_3}\cdots b^\dagger_{i_K}
\ket{0} \oplus \cdots
\label{cod39}
\eeq
This corresponds to the splitting of the ${\cal N}$-dimensional
representation of $SU({\cal N})$ as
\beq
\underline{{\cal N}} = (1, 1)\oplus (R^K_{K-1}, R^{N-K}_1) \oplus
\cdots \oplus (R^K_{K-l}, R^{N-K}_l) \oplus \cdots ( 1, R^{N-K}_K)
\label{cod40}
\eeq
where $R^K_{K-l}$ is the rank $(K-l)$ antisymmetric representation of
$SU(K)$ and $R^{N-K}_l$ is the rank $l$ antisymmetric representation of
$SU(N-K)$, with the corresponding dimensions
\beq
{\rm dim\,}[R^{K}_{K-l}] = {K! \over (K-l)!\, l! }, \hskip .3in
{\rm dim\,}[R^{N-K}_l ] = {(N-K)! \over l! \,(N-K-l)!}
\label{cod41}
\eeq
There is a $U(1)$ subgroup of $SU(N)$ which commutes with
$SU(K)$ and $SU(N-K)$. The representations given above carry
nonzero charges under this $U(1)$ subgroup. Since they are not relevant for
arriving at the parametrization we are seeking, we do not display them here.

The first term in the series in (\ref{cod40}) is given by
\beq
\ket{1, 1} \sim \epsilon_{i_1 i_2 \cdots i_K}\, b^\dagger_{i_1} b^\dagger_{i_2}\cdots b^\dagger_{i_K}\ket{0}
\label{cod42}
\eeq
It corresponds to the state designated as
$B = 1$ in writing $c_A = \U_{AB}\big\vert_{B=1} = \U_{A1}$.
This representation $(1,1)$ is invariant 
under $SU(K) \times SU(N-K)$.
Notice that the next term in the series in (\ref{cod39}) 
can be written as
\beq
\ket{a,i} \sim \epsilon_{i i_2 \cdots i_K}  b^\dagger_{a} b^\dagger_{i_2}\cdots b^\dagger_{i_K}
\ket{0} \sim 
(b^\dagger_{a} b _{i} ) \left( \epsilon_{i_1 i_2 \cdots i_K}b^\dagger_{i_1} b^\dagger_{i_2}\cdots b^\dagger_{i_K}
\ket{0}\right)  \sim t_{a i}  \, \ket{1,1}
\label{cod43}
\eeq
where $t_{a i} = b^\dagger_a b_i$ are generators of $SU(N)$ transformations.
In fact $t_{a i}$ and their hermitian conjugates 
define the elements of the Lie algebra orthogonal to
 $SU(K) \times SU(N- K) \times U(1) \subset SU(N)$, namely they correspond to 
translations in the coset $SU(N)/[SU(K) \times SU(N- K) \times U(1)]$.
This shows that, in a general linear combination of the states in (\ref{cod39}), we may
combine the first two terms as
 \beq
\ket{1,1} +
 z_{a i} t_{a i} \,\ket{1,1} 
 = (1 +  z_{a i} t_{a i}) \,\ket{1,1} 
 \approx g \, \ket{1,1}
 \label{cod44}
 \eeq
where $g \in SU(N)$. 

There are two key observations we can make at this point.
First, in (\ref{cod44}), we have the expansion of $g$ to linear order in $z_{ai}$.
This can be exponentiated with a general group element
$g \sim e^{i t_{ai} z_{ai}}$ acting on $\ket{1,1}$. 
Actually, we can take $g$ to be a general element of $SU(N)$.
This is because, if we split $g$ as 
$ e^{i t_{ai} z_{ai}} \, h$, $h \in H$, in a local parametrization,
$h$ acts as identity and drops out from consideration
since the state $\ket{1,1}$ is invariant under
$H= SU(K) \times SU(N- K)$.
Thus if we use a general $g$ in (\ref{cod44}), it is naturally
restricted to the coset $SU(N)/ H$, with the identification
$g \sim g h $, $h \in H$.

Secondly, we note that in using $g \sim e^{i t_{ai} z_{ai}}$, after exponentiation of
$1+ z_{ai} t_{ai}$, we encounter higher powers of $t_{ai}$ and $z_{ai}$.
These will produce
 higher terms in the series
(\ref{cod39}). However, 
although higher powers of $t\cdot z$ in the expansion of $g$
do lead to the higher representations of
$SU(N-K) \times SU(K) $ shown in (\ref{cod39}), no new parameters are introduced beyond $z_{ai}$. Therefore in a general linear combination of the
states in (\ref{cod39}), we must allow for arbitrary coefficients for the
representations $(R^K_{K-l}, R^{N-K}_l)$, $l \geq 2$.
Writing $g$ as a $K$-fold antisymmetric product of $U$'s as in 
(\ref{cod32}), we now see that a general parametrization of $c_A$ is given by
\beqar
c_A &=& \C \biggl( {1\over \sqrt{K!}}
U_{\alpha_1 i_1} U_{\alpha_2 i_2} \cdots U_{\alpha_K i_K}
\epsilon_{i_1 i_2 \cdots i_K}\nonumber\\
&&
+ \sum_{l = 2}^K {1\over \sqrt{ (K-l)! l!}}
U_{\alpha_1 a_1} \cdots U_{\alpha_l a_l} 
U_{\alpha_{l+1} j_{l+1}} \cdots U_{\alpha_K j_K}\,
(\phi_{a_1 a_2 \cdots a_l, j_1 \cdots j_l} \epsilon_{j_1 \cdots j_K})
\biggr)\nonumber\\
\C &=& {1 \over \sqrt{ 1+ \phi^*_{ab,ij} \phi_{ab, ij} + 
\phi^*_{abc,ijk} \phi_{abc,ijk} + \cdots}}\label{cod45}
\eeqar
We have chosen the normalization factors
to ensure that $c^*_A c_A =1$.
This is the parametrization which is suitable for explicitly carrying out
further reductions.

The parametrization (\ref{cod45}) has a natural interpretation in terms
of the physics of the unitary transformations.
The state $\ket{1, 1} \sim \epsilon_{i_1 i_2 \cdots i_K}\, b^\dagger_{i_1} b^\dagger_{i_2}\cdots b^\dagger_{i_K}\ket{0}$ is the many-particle state with
$K$ fermions that we start with.
The operator $t_{ai} = b^\dagger_a b_i$ represents moving one
particle from this occupied set to one of the unoccupied states
corresponding to $b^\dagger_a$. The quadratic term
in the expansion of
$e^{i t_{ai} z_{ai}}$, with $z_{ai} z_{bj} t_{ai} t_{bj} $ will correspond to
moving two particles from the occupied set to unoccupied states, but carried out one at a time.  Similar considerations apply to the higher powers.
Consider now the $l=2$ term in (\ref{cod45}), with
$\phi_{a_1 a_2 j_1 j_2} \epsilon_{j_1j_2 j_3\cdots j_K}$, followed by
a general unitary transformation of the single-particle Hilbert space
carried out by $U$.
This corresponds to moving a pair together (rather than one at a time)
from the occupied states to unoccupied ones and then allowing for
a general evolution at the level of single-particle Hilbert space.
Thus it can capture intrinsic two-particle correlations
which are not seen at the level of single particles.
Similarly, $l = 3, 4,\cdots $, will correspond to moving clusters of three
particles, four particles, etc. from occupied states to unoccupied ones,
carrying the physics of intrinsic many-body correlations.

The parametrization (\ref{cod45}) is still very general. From the mathematical
point of view, it is only a particular
choice of coordinates on $\mathbb{CP}^{{\cal N} -1}$, somewhat different from (\ref{cod30a}), so the use of
(\ref{cod45}) in (\ref{cod30}) will describe the full dynamics for the
$K$-fermion system, including multiparticle correlations as explained above.
If we make the approximation of setting 
$\phi = 0$, we have just the first term in (\ref{cod45}), so that
\beq
c_{\alpha_1 \alpha_2 \cdots \alpha_K} \approx {1\over \sqrt{K!}}
U_{\alpha_1 i_1} U_{\alpha_2 i_2} \cdots U_{\alpha_K i_K}
\epsilon_{i_1 i_2 \cdots i_K}
\label{cod33}
\eeq
This will describe the approximation where time-evolution
is restricted to $U(N)$ transformations of the single-particle Hilbert space.
This would be a good approximation if
intrinsic many-particle correlations are negligible.
One can still have interactions, but at the level of single-particle wave functions.
This is the Hartree-Fock approximation.\footnote{We want to emphasize that the action 
(\ref{cod30}) is to be used in the path-integral 
with the parametrization (\ref{cod45}). This would involve writing
the measure of integration, which is the volume element
for $\mathbb{CP}^{\N -1}$, in terms of the $U$'s and the $\phi$'s.
The approximation of setting $\phi =0$ really means that
many particle-correlations are negligible for
dynamical reasons, and so
we drop $\phi$'s from all terms, 
specifically the Hamiltonian, except for those involving
the time-derivative of the $\phi$'s.
The term with the time-derivatives will lead to states
characterized by the values of
$\phi$ (as with coherent states). These values are not changed by
time-evolution, since the reduced $H$ does not involve
$\phi$. The path-integral can still give a normalization
factor, which would be related to how much information is lost
(or entropy is generated) by the approximation.}

To see how this works out in detail, as 
an example, consider taking the $K$ states as the first $K$
states so that $i_1, i_2, \cdots, i_K = 1, 2, \cdots, K$.
With the factorized form (\ref{cod33}) we then find
\beqar
i c^*_A {\dot c}_A &=& i \epsilon_{i_1 i_2 \cdots i_K} U^*_{\alpha_1 i_1}
\cdots U^*_{\alpha_K i_K}  \left( {\dot U}_{\alpha_1 j_1} u_{\alpha_2 j_2} \cdots
+ U_{\alpha_1 j_1} {\dot U}_{\alpha_2 j_2} \cdots + \cdots
\right) \epsilon_{j_1 j_2 \cdots j_K}\nonumber\\
&=&i \sum_{i=1}^K U^*_{\alpha i} {\dot U}_{\alpha i}
\label{cod34}
\eeqar
The part of the Hamiltonian which is in the Lie algebra of $U(N)$
will act on the $c$'s of the form given in (\ref{cod33}) as
\beqar
c^*_A H_{AB} c_B &=&{1\over K!}\epsilon_{i_1 i_2 \cdots i_K} U^*_{\alpha_1 i_1}
\cdots U^*_{\alpha_K i_K} \biggl[ H_{\alpha_1 \beta_1}\,
\delta_{\alpha_2 \beta_2}  \cdots \delta_{\alpha_K \beta_K}
\nonumber\\
&&+ \delta_{\alpha_1 \beta_1}\,
H_{\alpha_2 \beta_2} \, \delta_{\alpha_3 \beta_3}\cdots \delta_{\alpha_K \beta_K} + \cdots\biggr]
U_{\beta_1 j_1} U_{\beta_2 j_2} \cdots U_{\beta_K j_K}
\epsilon_{j_1 j_2 \cdots j_K}\nonumber\\
&=& \sum_i U^*_{\alpha i} H_{\alpha \beta} U_{\beta i}
\label{cod35}
\eeqar
Notice that with these two terms the action $\S$ in (\ref{cod30}) 
has the form
\beq
\S = \int dt\, \Tr \left[ \rho_0 \left( i U^\dagger {\del U \over \del t}
- U^\dagger H U \right) \right] + \cdots
\label{cod36}
\eeq
where $\rho_0$ is the ``occupancy matrix". It is diagonal, has
entries equal to one for the occupied states ($K$ of them) and
zero for all other entries. It is not the density matrix for the many-particle system, but is related to it \cite{rosen1}-\cite{KN1}.

If there are multiparticle interaction terms in the Hamiltonian, we get additional terms.
For example, a two-particle interaction term will be of the form
$V_{\alpha_1 \alpha_2 \beta_1 \beta_2}$ in $H_{AB}$.
For such terms we get
\beqar
c^*_A H_{AB} c_B &=&{1\over K!}\epsilon_{i_1 i_2 \cdots i_K} U^*_{\alpha_1 i_1}
\cdots U^*_{\alpha_K i_K} \biggl[ V_{\alpha_1 \alpha_2 \beta_1 \beta_2} \otimes \delta_{\alpha_3 \beta_3} \otimes \cdots \nonumber\\
&&+ V_{\alpha_1 \alpha_3 \beta_1 \beta_3} \otimes \delta_{\alpha_2 \beta_2} \otimes \delta_{\alpha_4 \beta_4} \cdots
+  \delta_{\alpha_1 \beta_1}  \otimes V_{\alpha_2 \alpha_3 \beta_2 \beta_3} \otimes  \delta_{\alpha_4 \beta_4} \cdots\biggr]\nonumber\\
&&~\times U_{\beta_1 j_1} U_{\beta_2 j_2} \cdots U_{\beta_K j_K}
\epsilon_{j_1 j_2 \cdots j_K}\nonumber\\
&=&{1\over 2} U^*_{\alpha_1 i} U^*_{\alpha_2 j}
V_{\alpha_1 \alpha_2 \beta_1 \beta_2} \left(U_{\beta_1 i } U_{\beta_2 j}
- U_{\beta_1 j } U_{\beta_2 i} \right)
\label{cod37}
\eeqar
One can do a similar reduction for three-particle interactions, etc.
The action now becomes
\beqar
\S &=& \int dt\, \Tr \left[ \rho_0 \left( i U^\dagger {\del U \over \del t}
- U^\dagger H U \right) \right] \nonumber\\
&&- {1\over 2}\int dt\, U^*_{\alpha_1 i} U^*_{\alpha_2 j}
V_{\alpha_1 \alpha_2 \beta_1 \beta_2} \left(U_{\beta_1 i } U_{\beta_2 j}
- U_{\beta_1 j } U_{\beta_2 i} \right) + \cdots
\label{cod38}
\eeqar
We get a generalized coadjoint action for fermions which can accommodate
interparticle interactions.

It is also instructive to analyze the equations of motion which follow from
the action in (\ref{cod38}). For this purpose, it is useful to write the
two-particle interaction terms in $\S$ as
\beq
\S_{\rm int} = -{1\over 2} \int dt\, V_{\alpha_1 \alpha_2 \beta_1 \beta_2}
\left( \rho_{\beta_1 \alpha_1} \rho_{\beta_2\alpha_2} -
\rho_{ \beta_1 \alpha_2} \rho_{\beta_2 \alpha_1} \right)
\label{cod46}
\eeq
where $\rho_{ \beta\alpha} = \sum_i U^*_{\alpha_i} U_{\beta i}
= (U \rho_0 U^\dagger )_{\beta \alpha}$.
From variation of $U$, we then obtain the equations of motion as
\beqar
i {\del \rho \over \del t} &=& [H + {\tilde V}, \rho]\nonumber\\
{\tilde V}_{\alpha \beta} &=& \left( V_{\alpha \alpha_2 \beta \beta_2} -
V_{\alpha \alpha_2 \beta_2 \beta} \right) \rho_{\beta_2 \alpha_2}
\label{cod47}
\eeqar
\section{Symbols and star-products}
We will now consider further simplification using star-products and symbols.
Once we have approximated to the Hartree-Fock level, the space of interest is the single-particle Hilbert space $\H_N$.
We now consider a set of wave functions $\{ \psi_\alpha \}$
which constitute a basis for $\H_N$.
The set $\{ \psi_\alpha \}$ can be viewed as the result of quantization
of a classical phase space $M$.
The nature of this phase space clearly depends on the physical context.
For example, for quantum Hall effect on $S^2$, we can take
$\{ \psi_\alpha \}$ as the wave functions of the lowest Landau level
obtained by geometric quantization with holomorphic polarization
of the symplectic two-form
$\omega = n \Omega$, where $\Omega$ is the K\"ahler two-form for
$S^2$ and $n$ is an integer proportional to the magnetic flux on $S^2$
\cite{KN2}.
The wave functions $\{ \psi_\alpha\}$ are thus sections of a holomorphic line bundle
on $S^2$.
In higher dimensions, we can have a similar structure with a
phase space $M$ which is a
complex K\"ahler manifold \cite{{KN1},{KN2}}.
Notice that $\psi_\alpha$, being functions of the phase space coordinates
give an embedding of $M$ into $\mathbb{CP}^{N-1}$.
For sufficiently large $N$, there are many complex K\"ahler manifolds
$M$ (of complex dimension less than $(N-1)$)
which can serve as the phase space. 
Thus for sufficiently large $N$, a set of (holomorphic) wave functions
is available for defining the symbols as we do below.
(This is very much along the lines of the Kodaira embedding of a complex K\"ahler manifold into $\mathbb{CP}^k$, for sufficiently large $k$.)
As mentioned before, which
manifold we choose for $M$ is to be determined by the physics.
In situations where the given data is the space of single-particle wave functions $\H_N$, and we are unable to identify a manifold $M$
of dimension $< N-1$ whose quantization leads to $\H_N$,
we can even choose 
$M$ to be $\mathbb{CP}^{N-1}$ itself, with $\omega$ taken as
the K\"ahler form $\Omega$ of $\mathbb{CP}^{N-1}$, i.e., with $n =1$.

Once we have a set of wave functions $\{\psi_\alpha\}$ satisfying a suitable holomorphicity condition, we can define the symbol corresponding to an operator $A$ on $\H_N$ (or a matrix with matrix elements $A_{\alpha \beta}$)
by
\beq
(A) = {1\over \lambda} \sum_{\alpha, \beta}
\psi_\alpha \, A_{\alpha \beta} \, \psi^*_\beta
\label{cod48}
\eeq
where $\lambda $ is a number related to the normalization of the
$\psi_\alpha$.
The symbol $(A)$ is a function on $M$.
For example, for $S^2$, we can take
\beqar
\omega &=& - i {n \over 2} \Tr ( \sigma_3 \, g^{-1} dg\, g^{-1} dg )\nonumber\\
g&=& {1\over \sqrt{1+\bz z}} \left( \begin{matrix}
1& z\\ -\bz& 1\\ \end{matrix} \right) \, h\label{cod49}\\
h&=& e^{i {\sigma_3\over 2} \vf}, \hskip .3in
\sigma_3 = \left( \begin{matrix}
1& 0\\ 0& -1\\ \end{matrix} \right)
\nonumber
\eeqar
The group element $g$ modulo $h$ parametrizes $S^2 = SU(2)/U(1)$
in terms of complex coordinates $z$, $\bz$.
In this case, carrying out the geometric quantization
of $\omega$ in (\ref{cod49}), we find $N = n+1$ and
$\psi_\alpha$ are given in terms of the matrix elements
of $g$ in the rank $n$ representation as
\beqar
\psi_\alpha 
&=& \sqrt{n+1}\, \bra{j, \alpha} g \ket{j, - j}
= \sqrt{N} \bra{\alpha}g \ket{w}\nonumber\\
&\equiv& \sqrt{N} \,\D^{(j)}_{\alpha, w}(g)
\label{cod50}
\eeqar
where $j = {n\over 2}$ and $\ket{w} = \ket{j, -j}$.
These states are also obtained as the lowest Landau levels 
on $S^2$ for a Hamiltonian which is
proportional to the Laplacian (with covariant derivatives) on
$S^2$. The background magnetic field given by the
covariant derivatives is $B= (n/2 r^2)$ where $r$ is the radius of
$S^2$.

The symbol in (\ref{cod48}) is given by
\beq
(A) = \sum_{\alpha, \beta} 
\D^{(j)}_{\alpha, w}(g) \, A_{\alpha\beta} \D^{(j)*}_{\beta, w}(g)
= \bra{w} g^T A g^* \ket{w}
\label{cod51}
\eeq
corresponding to $\lambda = N$. ($g^T$ denotes the transpose of the matrix
$g$.)
The trace of a matrix $A$ can be written using the orthonormality of
$\{\psi_\alpha \}$ as
\beq
\Tr A = \sum_\alpha A_{\alpha\alpha} = \sum_{\alpha,\beta}
\int d\mu \, \psi_\alpha \, A_{\alpha\beta} \, \psi^*_\beta =
\lambda \int d\mu\, (A)
\label{cod52}
\eeq
where $d\mu$ is the volume element on $S^2$ given by the K\"ahler form
$\Omega$. 

The symbol for the product of two matrices is given by
the star-product of $(A)$ and $(B)$ which is a series involving derivatives 
of $(A)$ and $(B)$,
\beq
(AB) = (A)* (B) = (A) (B) + \cdots
\label{cod53}
\eeq
This result can be derived as follows. For the symbol of the product
$AB$ we find, from the definition in (\ref{cod51}),
\beqar
(AB) &=& \bra{w} g^T A B g^* \ket{w} \nonumber\\
&=&\bra{w} g^T A\, g^* g^T \, B g^* \ket{w} = \sum_k \bra{w} g^T A g^*\ket{k} \bra{k}  g^T \, B g^* \ket{w}\nonumber\\
&=& \bra{w} g^T A g^*\ket{w} \bra{w}  g^T \, B g^* \ket{w}
+ \bra{w} g^T A g^* J_+ \ket{w} {1\over n}\bra{w} J_- g^T \, B g^* \ket{w} + \cdots\nonumber\\
&=& (A) (B) - {1\over n} R_-(A) \, R_+(B)
+ {1\over 2 n (n-1)} R_-^2(A) \, R_+^2 (B)\nonumber\\
&&\hskip .1in
 + {\rm terms~with~higher~derivatives}\nonumber\\
 &=& (A) (B) + {1\over n} (1+ \bz z)^2 {\del (A) \over \del \bz}
 {\del (B) \over \del z} + \cdots \equiv (A)* (B),
\label{cod54}
\eeqar
where $J_\pm$ are raising and lowering
operators for $SU(2)$. The action of $J_+$ to the right on $g^*$,
and similarly $J_-$ to the left on $g^T$, can be obtained by the right translation differential operators on the group element
$g$ leading to the last two lines of (\ref{cod54}). 
This uses the fact that
the set $\{ \psi_\alpha\}$ are coherent states
for $SU(2)/U(1)$ obeying a holomorphicity condition
$R_- \psi_\alpha = 0$.
Explicitly, the differential operators $R_\pm$ are given by
\beqar
R_+ &=& e^{i \vf} \left[(1+ \bz z ) {\del \over \del z} - i \bz {\del \over \del \vf}
\right]
\nonumber\\
R_- &=& e^{-i \vf} \left[ - (1+ \bz z ) {\del \over \del \bz}
- i z {\del \over \del \vf} \right]
\label{cod55}
\eeqar
(The functions $(A)$, $(B)$ do not depend on $\vf$. However,
for multiple applications of $R_\pm$, there will be additional terms
arising from derivatives acting on $e^{\pm i \vf}$ factors.
These produce the needed Levi-Civita connection terms
in the covariant derivatives.
For example, $R_+ (B)$ is a component of a vector and hence must
have an appropriate connection term for subsequent application of
$R_\pm$.)
Although we used $S^2$ as an example,
the star-product for other complex K\"ahler manifolds can be derived along similar lines, see, for example, \cite{KN1}.
The result for the action and the Hamiltonian given below will hold 
for the case of $M$ being a general complex K\"ahler manifold
with the appropriate star-products.

We can now write the first term of the action in
(\ref{cod38}) in terms of symbols (i.e., functions on phase space)
and star-products as
\beq
\S = \lambda \int dt d\mu \left[  \rho_0 *\left(i U^\dagger* {\del U \over \del t}
- U^\dagger* H* U \right) \right] + \cdots
\label{cod56}
\eeq
(All quantities on the right hand side are the symbols of the corresponding operators; from now on we will drop the parentheses notation for symbols
to avoid clutter.)
It should be kept in mind that, if $U$ is of the form
$e^{i \Phi}$, the corresponding symbol is
\beq
U = 1 + i \Phi + {i^2 \over 2!} \Phi* \Phi + {i^3 \over 3!} \Phi*\Phi*\Phi + \cdots
\label{cod56a}
\eeq
$U$ will thus have terms with derivatives of $\Phi$ as well.
The action (\ref{cod56}) is what has been used in \cite{wadia}, \cite{sakita}
and \cite{KN1}
to analyze the dynamics of the quantum Hall droplet
or effectively bosonization around the Fermi surface which is the 
edge of the droplet.

Turning to the interaction terms in (\ref{cod46}), we can write
\beqar
V_{\alpha_1 \alpha_2 \beta_1 \beta_2}
\, \rho_{\beta_1 \alpha_1} \rho_{\beta_2\alpha_2}
&=&\lambda^2 \int d\mu(g) d\mu (g')
\left[\D_{\gamma ,w} (g) \rho_{\gamma \sigma} \D^*_{\sigma, \delta}(g)\right]
\left[\D_{\tau, w}(g') \rho_{\tau \mu}
\D^*_{\mu,\nu}(g')\right]\nonumber\\
&&\times
V_{\alpha_1 \alpha_2 \beta_1 \beta_2} \D_{\alpha_1, \delta}(g)\D_{\alpha_2,\nu}(g') \D^*_{\beta_1,w}(g) \D^*_{\beta_2, w}(g')\nonumber\\
&=&\lambda^2 \int d\mu(\bz, z) d\mu (\bz', z')
\rho(z,\bz)* \rho(z',\bz') *V(z,\bz; z',\bz')
\label{cod57}
\eeqar
where
\beq
V(z,\bz; z',\bz') = V_{\alpha_1 \alpha_2 \beta_1 \beta_2} \D_{\alpha_1, w}(g)\D_{\alpha_2,w}(g') \D^*_{\beta_1,w}(g) \D^*_{\beta_2, w}(g')
\label{cod58}
\eeq
The second interaction term in (\ref{cod46})
can be expressed in a similar way. To combine both, define
\beq
{\tilde V} (z,\bz; z',\bz')
= \bigl(V_{\alpha_1 \alpha_2 \beta_1 \beta_2} -
V_{\alpha_2 \alpha_1 \beta_1 \beta_2} \bigr)
\D_{\alpha_1, w}(g)\D_{\alpha_2,w}(g') 
\D^*_{\beta_1,w}(g) \D^*_{\beta_2, w}(g')
\label{cod59}
\eeq
This is the definition of the symbol corresponding to the two-particle
potential term. 
The interaction part of the action is then given by
\beq
\S_{\rm int} = - { \lambda^2\over 2} \int d\mu(\bz, z) d\mu (\bz', z')
\rho(z,\bz)* \rho(z',\bz') *{\tilde V(z,\bz; z',\bz')}
\label{cod60}
\eeq
It should be kept in mind that $\rho (z, \bz)$ is the symbol
$U* \rho_0 *U^\dagger$.
Further, it should be kept in mind that the star-product implies the
that there are terms with derivatives of $\rho$ and $V$ in
the interaction term.
Notice that the Hamiltonian in (\ref{cod56}) also involves
the combination $U \rho_0 U^\dagger$. We can therefore combine that
term with the interaction term as
\beq
H (\rho) = \lambda \int d \mu \, H*\rho + { \lambda^2\over 2} \int d\mu d\mu'\,
\rho(z,\bz)* \rho(z',\bz') *{\tilde V(z,\bz; z',\bz')}
\label{cod61}
\eeq
Since $H$ is given in terms of the symbol for $\rho$ (without
splitting it into $U$, $U^\dagger$ and $\rho_0$, one could think of
taking the expansion of the star-products in (\ref{cod61})
keeping $\rho$ as a function on the phase space.
It would be the analogue of the single-particle distribution in
classical Liouville formulation of phase space dynamics.
The action then takes the form
\beq
\S = \lambda \int dt d\mu\,  \left(\rho_0 *i U^\dagger* {\del U \over \del t}\right)
- \int dt \, H(\rho)
\label{cod62}
\eeq
Variants of this way of writing the action have appeared in
\cite{rosen1}-\cite{son}. The first term involving
the time-derivative of $U$ is sometimes referred to as the WZW term.
The fact that this term upon using the expansion of the star-product leads to the WZW action was first pointed out in \cite{{wadia},{sakita}}
and worked out in higher dimensional quantum Hall systems
(with both Abelian and nonabelian background fields) in
\cite{KN1}.

The star-products are defined in a series expansion, which is effectively
in inverse powers of an appropriate power of the
phase volume associated to
$\omega$. Of course, in practice, one has to truncate the
series at some finite order.
\section{Discussion}
In this note we gave a general formulation of the use of coadjoint orbits for
multifermion systems.
The dynamics of any quantum system is given in terms of a coadjoint orbit action
on $SU(\N) / U(\N -1)$ where $\N$ is the dimension of the Hilbert
space. (For infinite dimensional Hilbert spaces, one has to define
a suitable limit of $\N \rightarrow \infty$.)
The symplectic structure is defined by the K\"ahler form
on $SU(\N) / U(\N -1)$.
So far, it is an exact description.

If two-particle and many-particle correlations are negligible,
one can reduce this dynamics to unitary transformations
on the single-particle Hilbert space. 
The parametrization of the dynamical variables which naturally facilitates
this approximation is given in section 3.
The action then reduces to the 
coadjoint orbit action used in \cite{rosen1}-\cite{Kar}
as well as in most of the literature on bosonizations around the
Fermi surface.
This reduction is equivalent to the Hartree-Fock approximation.
Two-particle and many-particle interactions can be included, as expected,
 at the
level of a factorized form for the wave functions.
This is first step in a set of approximations or reductions.

The dynamics is still formulated in terms of
unitary operators or matrices.
As the second step, one can replace the operators (or matrices) 
by functions on the phase space, with star-products accounting 
for the noncommutative nature of the operator products.
In principle, there is no further approximation in this step.
However, one can then truncate the star-product expansion
at a finite order, which is the third step of approximation.
Explicit formulae for the Kac-Moody algebra of edge excitations
(\cite{wadia}-\cite{KN1}), for density correlations \cite{son}, etc.
have been obtained at this level.

Clearly the most interesting next step would be to consider
keeping intrinsic many-particle correlations, i.e., keeping some or all
of the $\phi_{ab, ij}, \, \phi_{abc,ijk}$, etc. in the parametrization (\ref{cod45}).
We hope to take up this question as well as the related question of how
much information is lost by the neglect of the $\phi$'s in future.

\bigskip
I thank Dimitra Karabali for a careful reading and for suggestions
on improving the presentation.
This work was supported in part by the U.S. National 
Science Foundation Grant No. PHY-2412479.

\end{document}